\documentclass[preprint, showpacs,preprintnumbers,amsmath,amssymb,nofootinbib]{revtex4}

 \usepackage{epsf}
 \usepackage{graphicx}

\begin{document}
 \newcommand{\be}[1]{\begin{equation}\label{#1}}
 \newcommand{\ee}{\end{equation}}
 \newcommand{\bea}{\begin{eqnarray}}
 \newcommand{\eea}{\end{eqnarray}}
 \def\disp{\displaystyle}

 \def\gsim{ \lower .75ex \hbox{$\sim$} \llap{\raise .27ex \hbox{$>$}} }
 \def\lsim{ \lower .75ex \hbox{$\sim$} \llap{\raise .27ex \hbox{$<$}} }

\title{\Large \bf The growth of matter perturbations in the $f(T)$ gravity}
\author{Xiangyun Fu$^{1}$, Puxun Wu$^{2}$ and Hongwei Yu$^{3,}$\footnote{Corresponding author:hwyu@hunnu.edu.cn}}

\address{$^1$ Institute of  Physics, Hunan University of Science and Technology, Xiangtan, Hunan 411201, China
\\
$^2$ Center for Nonlinear Science and Department of Physics,
Ningbo University, Ningbo, Zhejiang 315211, China\\
$^3$Department of Physics and Key Laboratory of Low Dimensional
Quantum Structures and Quantum Control of Ministry of Education,
Hunan Normal University, Changsha, Hunan 410081, China
}

 \begin{abstract}

In this paper, we study the growth index of matter density
perturbations for the power law model in $f(T)$ gravity. Using the
parametrization $\gamma(z)=\gamma_0+\gamma_1 {z\over 1+z}$ for the
growth index, which approximates the real evolution of $\gamma(z)$
very well, and the observational data of the growth factor, we find
that, at $1\sigma$ confidence level, the power law model in $f(T)$
gravity is consistent with the observations,  since the obtained
theoretical values of $\gamma_0$ and $\gamma_1$   are  in the
allowed region.
\end{abstract}

\pacs{95.36.+x, 98.80.Es}

 \maketitle
 \renewcommand{\baselinestretch}{1.5}

\section{Introduction}\label{sec1}
Cosmological data from a wide range of sources have indicated that
our Universe is undergoing an accelerating expansion~\cite{Sne, CMB,
SDSS}.  Basically, two kinds of alternative explanations have been
proposed for this unexpected observational phenomenon. One is the
dark energy with a sufficient negative pressure, which induces a
late-time accelerating cosmic expansion. Currently, there are many
candidates of dark energy, such as the cosmological constant,
quintessence, phantom, quintom, and so on. The other is the modified
gravity, which originates from the idea that the general relativity
is incorrect in the cosmic scale and therefore needs to be modified.
Examples of such theories are   the scalar-tensor
theory%~\cite{scalartensor}
, the $f(R)$ theory  %~\cite{frmodel}
and the
Dvali-Gabadadze-Porrati (DGP) braneworld scenario, %~\cite{r32,r33},
{\it et al.}.

Recently, a new interesting modified gravity by extending the
teleparallel theory~\cite{Einstein1930}, called $f(T)$ gravity, is
proposed to explain the present accelerating cosmic
expansion~\cite{Bengochea2009}. Since the teleparallel theory is
based upon the Weitzenb\"{o}ck connection rather than the
Levi-Civita one, it has no curvature but only torsion. In analogy to
the well-known $f(R)$ gravity obtained from extending the
Einstein-Hilbert action  to be an arbitrary function of $R$, the
$f(T)$ theory is built by generalizing the action  of the
teleparallel gravity to be $T+f(T)$~\cite{Bengochea2009,
Linder2010}. An important advantage of the $f(T)$ gravity is that
its field equations are second order as opposed to the fourth order
equations of the $f(R)$ gravity. So, it has spurred an increasing
deal of interest in the literatures~\cite{Yerzhanov2010,
Ferraro2007, Ferraro2008,Bamba2010, Wu2010a,Wu2010b, jbdent, rzheng,
Wu2010c} . For example, some concrete models are built, in
\cite{Bengochea2009, Linder2010, Wu2010c, Bamba2010},  to account
for the present cosmic expansion and the models with the phantom
divide line crossing are proposed in \cite{Wu2010c, Bamba2010}. The
observational constraints on model parameters are discussed in
\cite{Wu2010a} and the dynamical analysis for a general $f(T)$
theory is performed in~\cite{Wu2010b}.

 It has been pointed out~\cite{Linder2010, Wu2010a} that the
$f(T)$ theory can give the same background evolution as other
models, such as the $\Lambda$CDM and DGP. So, in order to
discriminate the $f(T)$ gravity from other models, one needs to
break the degeneracy of the background expansion history.
 An interesting approach  to differentiating  the modified gravity
and  dark energy  is to use  the growth function
$\delta(z)\equiv\delta\rho_m/\rho_m$ of the linear matter density
contrast~\cite{r1,r12,r13,r14,r15,r16,r17,aastarobinsky,
wangli,jnfry,r18,r19,r20,r22,r28,r29,r30,r31,weihao,bboisseau,gongyungui}.
While different models give the same late time expansion,  they may
produce different growth of matter
perturbations~\cite{aastarobinsky}. To discriminate different models
with the matter perturbation, usually, the growth factor
$g\equiv\frac{d\ln\delta}{d\ln a}$ is used and it can be
parameterized as~\cite{wangli,jnfry}
 \begin{equation}
 \label{fommegam}
 g\equiv\frac{d\ln\delta}{d\ln a}=\Omega_m^\gamma,
 \end{equation}
where $\gamma$ is the growth index and  $\Omega_m$ is the fractional
energy density of matter. This approach has been explored in many
works~\cite{r18,r19,r20,r22,r28,r29,r30,r31,weihao,bboisseau,gongyungui},
and it is found that different models may give different values of
the growth index if a constant $\gamma$ is considered. For example,
$\gamma_\infty\simeq 0.5454$~\cite{r18,r19} for  the  $\Lambda$CDM
model and $\gamma_\infty \approx0.6875$~\cite{r18,weihao} for the
flat DGP model. Therefore, in principle, one can distinguish the
modify gravity from dark energy  with the observational data on the
growth factor.

Recently, the authors in Refs.~\cite{jbdent,rzheng} have discussed
the matter perturbations in the $f(T)$ gravity and  found that,
despite the completely indistinguishable background behavior, the
growth of  matter density perturbation of different models can be
different. In this paper,  we plan to investigate the growth index
of matter density perturbations in the $f(T)$ gravity, and then by
using a feasible parametrization of  the growth index to compare the
growth function with the observational data.  Our results show that
the $f(T)$ gravity is consistent with the observations.

%%%%%%%%%%%%%%%%%%%%%%%%%%%%
\section{The $f(T)$ theory}
In this section, we give a brief review of the $f(T)$ gravity.  The
torsion scalar $T$ in the action of the teleparallel gravity is
defined as
 \begin{eqnarray}\label{t1}
 T\equiv S^{\;\;\mu\nu}_\sigma T^\sigma_{\;\;\mu\nu}\;,
 \end{eqnarray}
where
\begin{eqnarray}
S^{\;\;\mu\nu}_\sigma\equiv\frac{1}{2}(K^{\mu\nu}_{\;\;\;\;\sigma}+\delta^\mu_\sigma T^{\alpha \nu}_{\;\;\;\;\alpha}-\delta^\nu_\sigma T^{\alpha \mu}_{\;\;\;\;\alpha})\;,
\end{eqnarray}
and $T^\sigma_{\;\;\mu\nu}$ is the torsion tensor
\begin{eqnarray}
T^\sigma_{\;\;\mu\nu}\equiv e^\sigma_A (\partial_\mu e^A_\nu- \partial_\nu e^A_\mu )\;.
\end{eqnarray}
Here $e^A_\mu$ is the orthonormal tetrad component, where $A$ is an
index running over $0,1,2,3$ for the tangent space of the manifold,
while $\mu$, also running over   $0,1,2,3$, is the coordinate index
on the manifold.  $K^{\mu\nu}_{\;\;\;\;\sigma}$ is the contorsion
tensor given by
\begin{eqnarray}
 K^{\mu\nu}_{\;\;\;\;\sigma}=-\frac{1}{2}(T^{\mu\nu}_{\;\;\;\;\sigma}-T^{\nu\mu}_{\;\;\;\;\sigma}-T_{\sigma}^{\;\;\mu\nu})\;.
 \end{eqnarray}
Similar to $f(R)$ gravity, the $f(T)$ theory is obtained by
extending the action of teleparallel gravity to be $T + f(T)$ to
explain the late time accelerating cosmic  expansion with no need of
an exotic dark energy.

Assuming a flat homogeneous and isotropic Friedmann-Robertson-Walker
universe described by the metric
\begin{eqnarray}
ds^2=dt^2-a^2(t)\delta_{ij}dx^idx^j\;,
\end{eqnarray}
where $a$ is the scale factor, one has, from Eq.~(\ref{t1}),
\begin{eqnarray}
T=-6 H^2\;,
\end{eqnarray}
with $H=\dot{a}/a$ being the Hubble parameter. Thus, in $f(T)$
gravity, the cosmic background equation can be written as
\begin{eqnarray}\label{habble}
H^2=\frac{8\pi G}{3}\rho-\frac{f}{6}-2H^2f_{T}\;,
\end{eqnarray}
\begin{eqnarray} \label{hubble2}
\dot{H}=-{1\over 4}\frac{6H^2+f+12H^2f_{T}}{1+f_{T}-12H^2f_{T T}}\;,
\end{eqnarray}
with $f_T\equiv df/dT$. Here we assume that the energy component in
our universe is only the matter with radiation neglected.
Apparently, the last two terms in the right hand side of
Eq.~(\ref{habble}) can be regarded as an effective dark energy.
Then, its effective energy density and equation of state can be
expressed, respectively, as
\begin{eqnarray}
\rho_{eff}=\frac{1}{16\pi G}(-f + 2Tf_{T})
\end{eqnarray}
\begin{eqnarray}
w_{eff}=-\frac{f/T-f_{T}+2Tf_{TT}}{(1+f_{T}+2Tf_{TT})(f/T-2f_{T})}\;.
\end{eqnarray}

In order to explain the present accelerating cosmic expansion, some
$f(T)$ models are proposed in Refs.~\cite{Bengochea2009, Linder2010,
Wu2010c, Bamba2010}. In this paper, we only consider a power law
model~\cite{Bengochea2009, Linder2010}
\begin{eqnarray}
f(T)=\alpha (-T)^n=\alpha(6H^2)^n\;,
\end{eqnarray}
with
\begin{eqnarray}\label{alpha}
\alpha=(6H_0^2)^{1-n}\frac{1-\Omega_{m0}}{2n-1}\;,
\end{eqnarray}
where $\Omega_{m0}=\frac{8\pi G\, \rho(0)}{3H_0^2}$ is the
dimensionless matter density parameter today.  Substituting above
two expressions into  Eq.~(8) and defining $E^2=H^2/H^2_0$, one has
\begin{eqnarray}\label{Ezd}
E^2=\Omega_{m0}(1+z)^3+(1-\Omega_{m0}) E^{2n}\;.
\end{eqnarray}
The reason to consider the power model in our paper is that  it has
the same background evolution equation as some phenomenological
models~\cite{Dvali2003, Chung2000} and  it reduces to the
$\Lambda$CDM model when $n=0$, and to the DGP model~\cite{Dvali2000}
when $n=1/2$.  In addition, it has a smaller $\chi^2_{Min}$ value
than the $\Lambda$CDM when fitting the recent observations such as
the Type Ia Supernova (Sne Ia), the baryonic acoustic oscillation
(BAO) and the Cosmic Microwave Background (CMB)
radiation~\cite{Wu2010a}.
 Let us note that,
in order to be consistent with the present observational results, it
is required that $|n|\ll 1$~\cite{Bengochea2009, Linder2010,
Wu2010a}.

\section{Growth index of $f(T)$ model}\label{sec2}
To the linear order of matter density perturbations,  the growth
function $\delta(z)$ at scale much smaller than the Hubble radius
satisfies the following equation~\cite{wangli}
\begin{equation}
\label{denpert} \ddot{\delta}+2H\dot{\delta}-4\pi
G_{eff}\,\rho_m\delta=0,
\end{equation}
where  $G_{eff}$ is the effective Newton's constant and the dot
denotes the derivative with respect to time $t$. In the general
relativity,  $G_{eff}= G_N$, where $G_N$ is the Newton's constant.
Defining the growth factor $g\equiv d\ln\delta/d\ln a$, one can
obtain
\begin{equation}
\label{grwthfeq1} {d\; g\over d\ln
a}+g^2+\bigg(\frac{\dot{H}}{H^2}+2
\bigg)g=\frac{3}{2}\frac{G_{eff}}{G_{N}}\Omega_m.
\end{equation}
In general, the analytical solution of above equation is very
difficult to obtain, and thus, we need to resort to the numerical
methods.

In the $f(T)$ gravity,  $G_{eff}$ can be expressed as~\cite{rzheng}
 \be{eq12}
 {G_{eff}}= {G_{N}\over 1+f_T}.
 \ee
So, the growth factor  satisfies the following equation:
 \be{eq16}
{d\; g\over d\ln
a}+g^2+\bigg(\frac{\dot{H}}{H^2}+2
\bigg)g
 =\frac{3}{2}{1\over 1+f_T}\Omega_m\,.
 \ee
Using
 \be{eq17}
\frac{\dot{H}}{H^2}=-\frac{3}{2}
 \frac{1+f/6H^2+2f_T}{1+f_T-12H^2f_{TT}}\,,
 \ee
 and Eq.~(\ref{alpha}), we get
 \begin{eqnarray}
 \label{eqg2}
 {d\; g\over d\ln
a}+g^2
 +g\left[2-{3\over2}\frac{1-E^{2n-2}(1-\Omega_{m0})}{1-nE^{2n-2}(1-\Omega_{m0})}\right]
 =\frac{3}{2}\frac{\Omega_{m}}{1-{n(1-\Omega_{m0})E^{2n-2}\over 2n-1}},
 \end{eqnarray}
 where $\Omega_m=\Omega_{m0}E^{-2}(1+z)^{-3}$ and $E$ satisfies
  \begin{eqnarray}
 \label{hubble0}
 {d E^2\over d\ln a}=\frac{-3E^2+3E^{2n}(1-\Omega_{m0})}{1-n E^{2n-2}(1-\Omega_{m0})}\,.
 \end{eqnarray}
Thus, given the values of $\Omega_{m0}$ and $n$, the value of $g$
can be obtained by solving Eqs.~(\ref{eqg2}) and (\ref{hubble0})
numerically with the initial condition $g=1$ at
$z\rightarrow\infty$. Then, using Eq.~(\ref{fommegam}), we can get
the value of growth index. The results are shown in
Fig.~(\ref{fommegam}). From the left panel of this figure, we can
see that, when $n=0$, our result reduces to that in the
$\Lambda$CDM. As expected, the growth factor $g$ in $f(T)$ gravity
($n\neq0$) grows slower than that in general relativity, which is
the same as that obtained  in Refs.~\cite{rzheng,jbdent},  because a
weak  effective Newton gravity (see Eq.~(\ref{eq12})) is obtained.
 The right panel shows that the growth index $\gamma$ in $f(T)$
gravity is different from that in the $\Lambda$CDM  at both low
redshifts and high redshifts.  This is caused by the fact that the
$f(T)$ gravity and $\Lambda$CDM give  different background
evolutions  at low redshifts. However, $\gamma$ is mainly determined
by the low redshift evolution,  since, at high redshifts
$\Omega_{m}\sim 1$, different values of $\gamma$ may give the same
result for $g$.  When $\Omega_{m0}=0.272$, the values of growth
index $\gamma$ approach to $0.58$ and $0.71$ with $z\rightarrow
\infty$ for $n=0.1$ and $0.25$, respectively. These results are
different from that obtained from the $\Lambda$CDM and DGP models
where the values are $\gamma_\infty\approx 0.5454$~\cite{r18,r19}
and $\gamma_\infty \approx0.6875$~\cite{r18,weihao}, respectively.
This feature of $\gamma_\infty$ provides a distinctive signature for
$f(T)$ as opposed to the $\Lambda$CDM and DGP  models. So,  one can
distinguish the $f(T)$gravity from other cosmological models  by the
growth index of matter density perturbations.
\begin{center}
 \begin{figure}[htbp]
  \label{fggrowthr}
 \centering
 \includegraphics[width=0.45\textwidth]{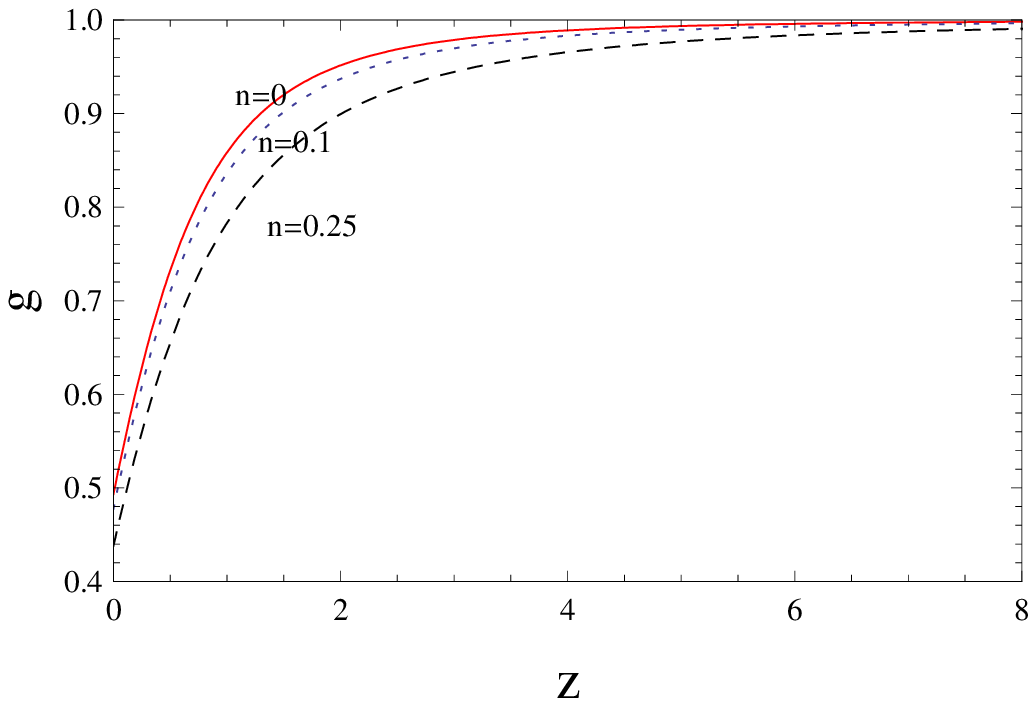}
 \includegraphics[width=0.45\textwidth]{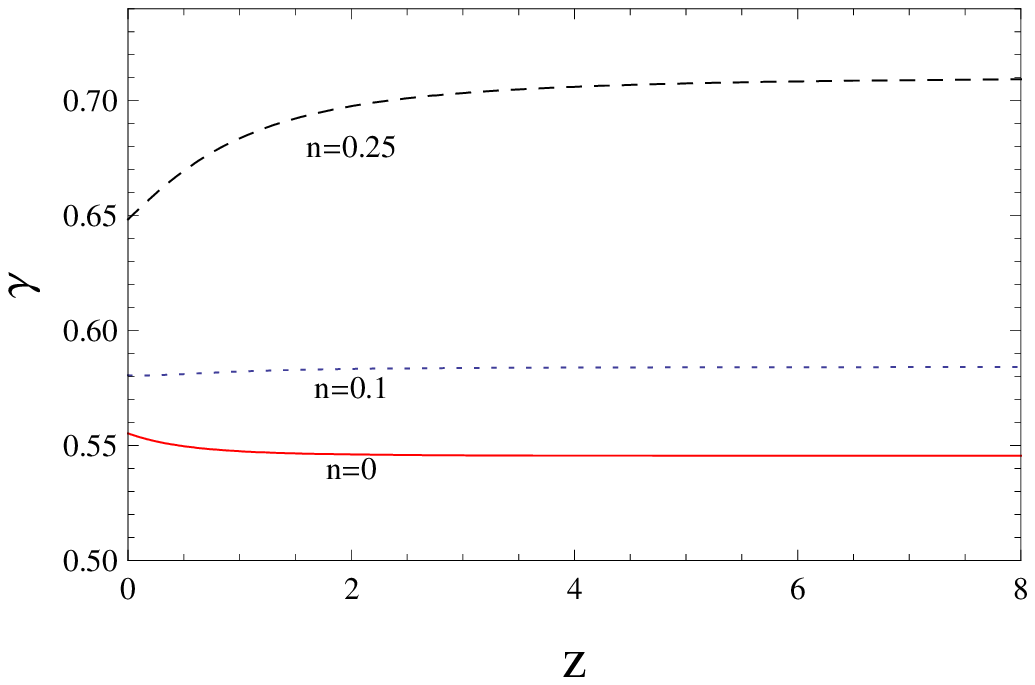}
 \caption{\label{figgamma0}
 while $\Omega_{m0}=0.272$, $g$ and $\gamma$ are  displayed as a function of $z$ for $f(T)$ gravity respectively. }
 \end{figure}
 \end{center}

 \section{Growth index parametrization and Observational constraints}
From the right panel of Fig.~(\ref{figgamma0}), it is easy to see
that, for any $n$, $\gamma$ is not a constant, especially in the
redshifts region $(z<2)$ where some observational data points are
obtained. So, it is unreliable to discriminate different models with
these observational data if  the growth index is treated as a
constant. The growth index $\gamma$ should be a function of $z$ and
we may parameterize it. Since Eq.~(\ref{eqg2}) can be reexpressed as
\begin{eqnarray}
 \label{eqg3}
 &&{3\Omega_{m}(n-1)E^{2n-2}(1-\Omega_{m0})\over 1-n E^{2n-2}(1-\Omega_{m0})}{dg \over d\Omega_m}
 +g^{ 2}
 +g\left[2-{3\over2}\frac{1-E^{2n-2}(1-\Omega_{m0})}{1-nE^{2n-2}(1-\Omega_{m0})}\right]\nonumber\\
 & =&\frac{3}{2}\frac{\Omega_{m}}{1-{n(1-\Omega_{m0})E^{2n-2}\over
 2n-1}},
 \end{eqnarray}
substituting  Eq.~(\ref{fommegam}) into the above expression, we
obtain an equation of $\gamma (z)$
 \begin{eqnarray}
 \label{gammaprime}
  &&-(1+z)\ln{\Omega_m}\gamma^\prime+\Omega_m^\gamma+(2-3\gamma)-{3\over2}(1-2\gamma)\frac{1-E^{2n-2}(1-\Omega_{m0})}{1-nE^{2n-2}(1-\Omega_{m0})}
  \nonumber\\&=&\frac{3}{2}\frac{\Omega_{m}^{1-\gamma}}{1-{n(1-\Omega_{m0})E^{2n-2}\over 2n-1}}\,.
 \end{eqnarray}

\begin{center}
 \begin{figure}[htbp]
\centering
 \includegraphics[width=0.45\textwidth]{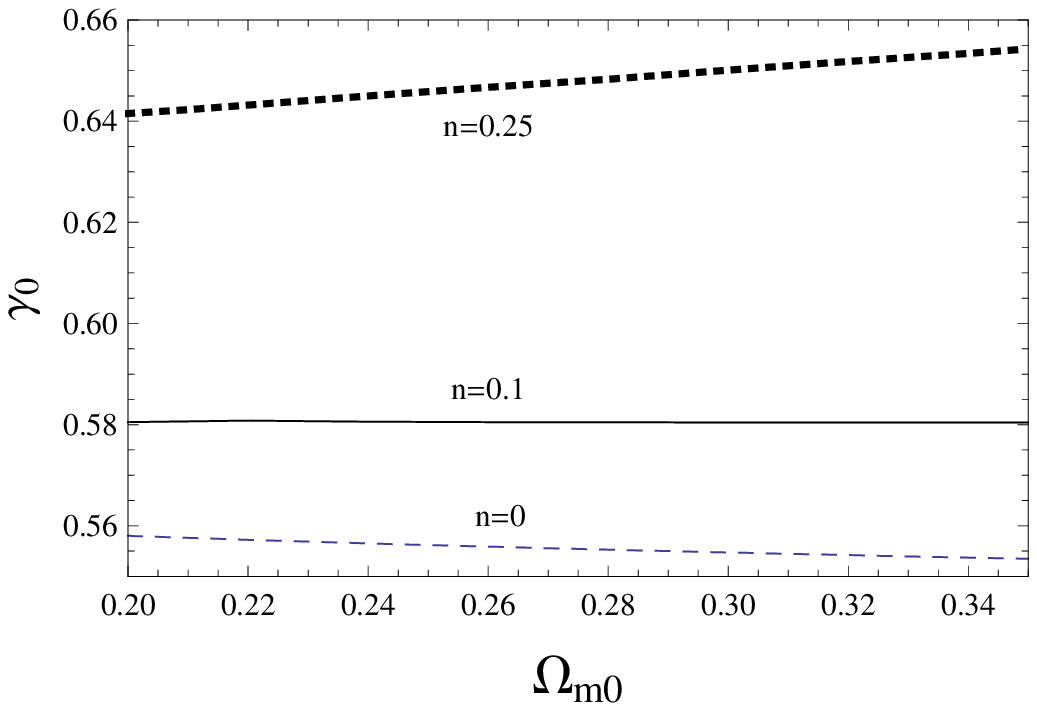}
 \includegraphics[width=0.45\textwidth]{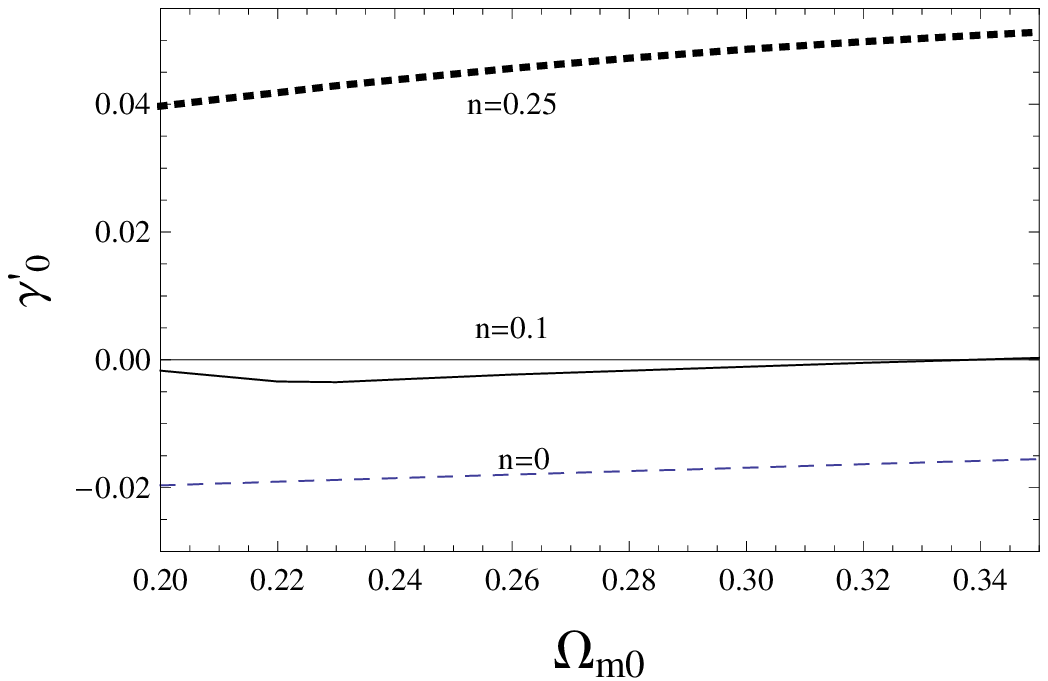}
 \caption{\label{gamma0gamm01}  $\gamma_0$ and
  $\gamma_1$ are displayed as a function of $\Omega_{m0}$ for $n=0, 0.1, 0.25$ respectively.}
 \end{figure}
 \end{center}

 Here, we
consider a parametrization form of $\gamma(z)$ proposed in
Ref.~\cite{wupfx}
\begin{equation}
\label{gammazpa} \gamma(z)=\gamma_0+\gamma_1{z\over 1+z},
\end{equation}
which gives a very good approximation of $\gamma(z)$ for the $w$CDM
and DGP models. The error is below $0.03\%$ for the $\Lambda$CDM
model and $0.18\%$ for the DGP model for all redshifts when
$\Omega_{m0} = 0.27$. In $f(T)$ gravity, for any given values of $n$
and $\Omega_{m0}$, we can obtain the value of $\gamma_0$ through the
value of $g_0$ determined from solving Eq.~(\ref{eqg3}) numerically.
Substituting Eq.~(\ref{gammazpa}) into Eq.~(\ref{gammaprime}), we
can get the expression of $\gamma_1$ which is a function of
redshifts $z$. For simplicity, we take the value of $\gamma_1$ at
$z=0$,
 \begin{eqnarray}
 \label{gamma0prime-1}
  \gamma_1&=&(\ln\Omega_{m,0}^{-1})^{-1}\bigg[-\Omega_{m0}^{\gamma_0}+{3\over
  2}{\Omega_{m0}^{1-\gamma_0}\over 1-{n(1-\Omega_{m0})\over 2n-1}}-(2-3\gamma_0)+{3\over 2}{\Omega_{m0}(1-2\gamma_0)\over 1-n(1-\Omega_{m0})}\bigg]\,.
 \end{eqnarray}
Thus, we obtain the possible regions of $\gamma_0$ and $\gamma_1$
for $0.20\leq\Omega_{m0}\leq 0.35$. The results are shown in
Fig.~(\ref{gamma0gamm01}). We find,  for a given value of
$\Omega_{m0}$, that the values of $\gamma_0$ and $\gamma_1$ in
$f(T)$ gravity ($n\neq 0$) are larger than those in the $\Lambda$CDM
model ($n=0$), and a larger value of $n$ gives  larger values of
$\gamma_0$ and $\gamma_1$. The reason for these is  that the
effective gravity in $f(T)$ theory  is weaker than that in general
relative, and a larger $n$ leads to a weaker effective gravity.
These features provide distinctive signatures for $f(T)$ gravity  as
opposed to the $\Lambda$CDM model. Therefore, in principle, we can
discriminate the $f(T)$ model from the $\Lambda$CDM model merely
through the values of $\gamma_0$ and $\gamma_1$ if one can obtain
their accurate values from the observation data.

\begin{center}
 \begin{figure}[htbp]
 \centering
 \includegraphics[width=0.6\textwidth]{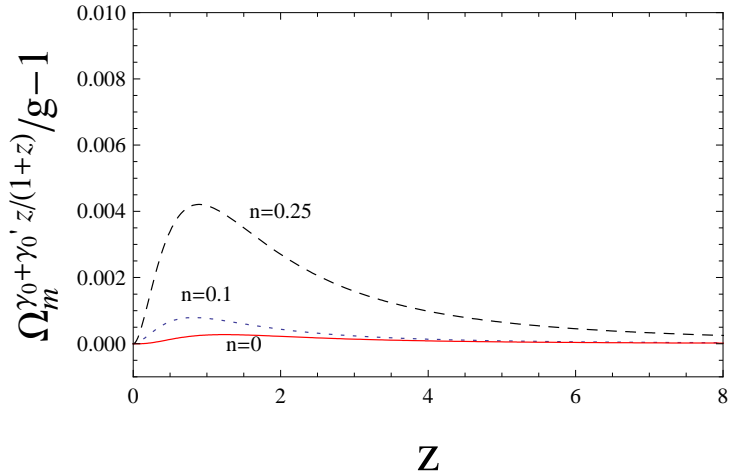}
\caption{\label{parametrization}  The relative difference between
the growth factor $g$ and $\Omega_m^\gamma$ with
$\gamma=\gamma_0+\gamma_1\frac{z}{1+z}$ and $\Omega_{m0}=0.272$. The
dashed, dotted and solid curves show the results of $n=0.25, 0.1,
0$, respectively.}
 \end{figure}
 \end{center}

Before studying the observational constraints on $\gamma_0$ and
$\gamma_1$, we need to examine how well the $\Omega_m^{\gamma(z)}$
with $\gamma(z)$ taking the parametrization in Eq.~(\ref{gammazpa})
approximates the growth factor $g$. Numerical results are shown in
Fig.~(\ref{parametrization}) with $\Omega_{m0}=0.272$. From this
figure, we see that  the error is below $0.45\%$ for $n=0.25$ and
below $0.1\%$ for $n=0.1$.  So, $\Omega_m^{\gamma_0+\gamma_1{z\over
1+z}}$ approximates the growth factor $g$ very well both at low and
high redshifts in $f(T)$ theory, and  we can use   all data points
to constrain this parametrization.

 In order to obtain the observational constraints on $\gamma_0$ and
$\gamma_1$, we first need to determine the value of $\Omega_{m0}$
and $n$ from the observations. Here we use the results obtained from
the combination of the latest Union2 Type Ia Supernova (Sne Ia) set,
the BAO from the SDSS data and the Cosmic Microwave Background (CMB)
radiation~\cite{Wu2010a}. At the $95\%$ confidence level,
$\Omega_{m0}=0.272_{-0.032}^{+0.036}$, $n=0.04_{-0.33}^{+0.22}$ for
the power law model. With these best fit values and
Eqs.~(\ref{eqg2},  \ref{gamma0prime-1}), we obtain that the
corresponding theoretical values of $\gamma_0$ and $\gamma_1$ are
$\gamma_0=0.564$ and $\gamma_1=-0.0123$.

\begin{center}
 \begin{figure}[htbp]
 \centering
 \includegraphics[width=0.45\textwidth]{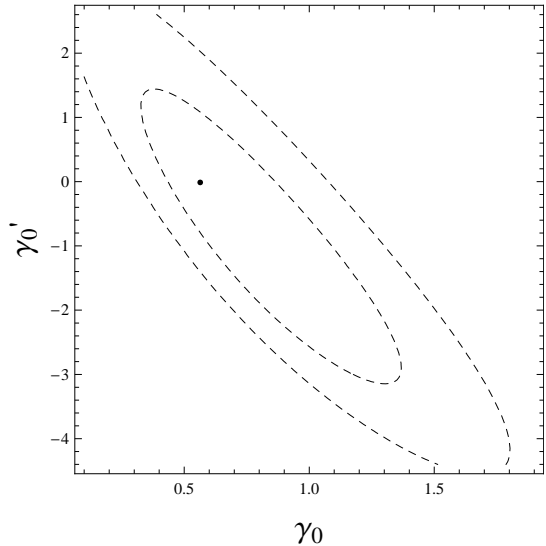}
 \caption{\label{constraints}  The $1 \sigma$ and $2\sigma$ contours of $\gamma_0$ and
  $\gamma_1$ by fitting the power law  model in $f(T)$ gravity  to the growth rate data.
 The point denotes  the theoretical values of $\gamma_0$ and
  $\gamma_1$ with the $\Omega_{m0}, n$ taking the best fit values.}
 \end{figure}
 \end{center}

Now we discuss the observational constraints on $\gamma_0$ and
$\gamma_1$ from the growth factor data. Here $12$ data points given
in Table~\ref{fzdata} are used.  Let us note that although the data
given in Refs.~\cite{viel1,viel2} are measured without `any' bias,
other data points are obtained by assuming a flat $\Lambda$CDM model
with $\Omega_{m0}$ taking a specific value, for example,
$\Omega_{m0}=0.25$ or $0.30$.  So, caution must be exercised when
using these data.  With this caveat in mind, it may still be
worthwhile to apply the data to fit the models~\cite{Gong2008,
Gong2009, weihao}. With the best fit values of $\Omega_{m0}$ and
$n$, we can obtain the constraints on $\gamma_0$ and $\gamma_1$ from
the observations by using the following equation
\begin{equation}
\label{fzchi}
\chi^2_g=\sum_{i=1}^{12}\frac{[g_{obs}(z_i)-\Omega_m^{\gamma_0+\gamma_1
z_i/(1+z_i)}]^2}{\sigma_{gi}^2},
\end{equation}
where $\sigma_{gi}$ is the $1 \sigma$ uncertainty of the $g(z_i)$
data. We find that the best fit values are $\gamma_0=0.809$ and
$\gamma_1=-0.942$. The allowed regions at $1$ and $2\sigma$
confidence levels are shown in Fig.(~\ref{constraints}),  from
which, one can see that the power law model in $f(T)$ gravity is
consistent with the observations, since the theoretical values of
$\gamma_0$ and $\gamma_1$ obtained by using the best fit values of
$\Omega_{m0}$ and $n$ for the power law model are in the allowed
region at $1 \sigma$ confidence level.

 \begin{table}[htp]\center
\begin{tabular}{|c|c|c|}
\hline
\ \ \ \ \ \ \ \ \   $z$\ \ \ \ \ \ \ \ \   & \ \ \ \ \ \ \  $g_{obs}$\ \ \ \ \ \ \  &\ \ References \\
\hline
$0.15$ & $0.49\pm 0.1$ & \cite{guzzo,guzzo2} \\
\hline
$0.35$ & $0.7\pm 0.18$ & \cite{tegmark} \\
\hline
$0.55$ & $0.75\pm 0.18$ & \cite{ross} \\
\hline
$0.77$ & $0.91\pm 0.36$ & \cite{guzzo,guzzo2} \\
\hline
$1.4$ & $0.9\pm 0.24$ & \cite{angela} \\
\hline
$3.0$ & $1.46\pm 0.29$ & \cite{mcdonald} \\
\hline
$2.125-2.72$ & $0.74\pm 0.24$ & \cite{viel1} \\
\hline
$2.2-3$ & $0.99\pm 1.16$ & \cite{viel2}\\
\hline
$2.4-3.2$ & $1.13\pm 1.07$ & \cite{viel2}\\
\hline
$2.6-3.4$ & $1.66\pm 1.35$ & \cite{viel2} \\
\hline
$2.8-3.6$ & $1.43\pm 1.34$ & \cite{viel2} \\
\hline
$3-3.8$ & $1.3\pm 1.5$ & \cite{viel2}\\
\hline
\end{tabular}
\caption{The summary of the observational data on the growth factor
$g$.} \label{fzdata}
\end{table}

\section{Conclusion}\label{sec4}
In this paper, we study in detail the growth index  of matter
density perturbations for  the power law model in $f(T)$ gravity.
Using a parametrization $\gamma(z)=\gamma_0+\gamma_1 {z\over 1+z}$
for the growth index $\gamma(z)$, which gives a very good
approximation of $\gamma(z)$, we find that the value of $\gamma_0$
and $\gamma_1$ in $f(T)$ gravity are larger than those in the
$\Lambda$CDM model, and a larger value of $n$ gives  larger values
of $\gamma_0$ and $\gamma_1$.  This feature may provide a
 signature for $f(T)$ gravity distinctive from the other models, such
as the $\Lambda$CDM and DGP.  Finally,  we discuss the constraints
on $\gamma_0$ and $\gamma_1$ from the observational growth factor
data and find that, at $1\sigma$ confidence level, the power law
model in $f(T)$ gravity is consistent with the observations since
the theoretical values of $\gamma_0$ and $\gamma_1$ obtained by
using the best fit values of $\Omega_{m0}$ and $n$  are in the
allowed region.

\section*{Acknowledgments}
This work was supported in part by the National Natural Science
Foundation of China under Grants Nos. 10935013 and 11075083,
Zhejiang Provincial Natural Science Foundation of China under Grant
Nos. Z6100077 and R6110518, the FANEDD under Grant No. 200922, the
National Basic Research Program of China under Grant No.
2010CB832803, the NCET under Grant No. 09-0144, and the PCSIRT under
Grant No. IRT0964.

\end{document}